\documentclass[apj]{emulateapj}
\slugcomment{{\sc Accepted to ApJL:} January 30, 2012}

\usepackage{graphicx}
\usepackage{epstopdf}
\usepackage{stmaryrd}
\usepackage{amsmath}
\usepackage{ccaption}
\usepackage{multirow}
\usepackage{bm}
\usepackage{nicefrac}
\usepackage{comment}
\usepackage{textcomp}
\usepackage{color}
\definecolor{red}{rgb}{1,0,0}

\bibpunct[; ]{(}{)}{,}{a}{}{,}

\newcommand{\msun}{$M_{\odot}$}
\newcommand{\mstar}{$M_{\star}$}
\newcommand{\lsun}{$L_{\odot}$}

\newcommand{\bdm}{\begin{displaymath}}
\newcommand{\edm}{\end{displaymath}}
\newcommand{\beq}{\begin{equation}}
\newcommand{\eeq}{\end{equation}}
\newcommand{\bit}{\begin{itemize}}
\newcommand{\eit}{\end{itemize}}
\newcommand{\ben}{\begin{enumerate}}
\newcommand{\een}{\end{enumerate}}
\newcommand{\bfi}{\begin{figure}[htb]}
\newcommand{\bpfi}{\begin{figure}[p]}

\newcommand{\lir}{$L_{\rm IR}$}

\shorttitle{IR LFs at $z$\,$<$\,2: main-sequence and starburst contributions}
\shortauthors{Sargent et al.}

\begin{document}


\title{The contribution of starbursts and normal galaxies to infrared luminosity functions at $z$\,$<$\,2}

\author{M.~T. Sargent\altaffilmark{1, $\star$},
M. B\'ethermin\altaffilmark{1},
E. Daddi\altaffilmark{1},
D. Elbaz\altaffilmark{1}}

\altaffiltext{$\star$}{~E-mail: \texttt{mark.sargent@cea.fr}}

\altaffiltext{1}{~CEA Saclay, DSM/Irfu/Service d'Astrophysique, Orme des Merisiers, F-91191 Gif-sur-Yvette Cedex, France}

\begin{abstract}
We present a parameterless approach to predict the shape of the infrared (IR) luminosity function (LF) at redshifts $z$\,$\leq$\,2. It requires no tuning and relies on only three observables: (1) the redshift evolution of the stellar mass function for star-forming galaxies, (2) the evolution of the specific star formation rate (sSFR) of main-sequence galaxies, and (3) the double-Gaussian decomposition of the sSFR distribution at fixed stellar mass into a contribution (assumed redshift- and mass-invariant) from main-sequence and starburst activity.\\
This self-consistent and simple framework provides a powerful tool for predicting cosmological observables: observed IR LFs are successfully matched at all $z$\,$\leq$\,2, suggesting a constant or only weakly redshift-dependent contribution (8\%--14\%) of starbursts to the star formation rate density. We separate the contributions of main-sequence and starburst activity to the global IR LF at all redshifts. The luminosity threshold above which the starburst component dominates the IR LF rises from log(\lir/\lsun)\,=\,11.4 to 12.8 over 0\,$<$\,$z$\,$<$\,2, reflecting our assumed (1+$z$)$^{2.8}$-evolution of sSFR in main-sequence galaxies.
\end{abstract}


\section{Introduction}
\label{sect:intro}

Determining the galaxy stellar mass function (MF) and the star formation rate (SFR) distributions -- e.g. the infrared (IR) luminosity function (LF) -- is among the foremost goals of extragalactic surveys targeting (and linking) nearby \citep[e.g.][]{bell03, baldry11, sanders03, goto11, bothwell11} and distant galaxies \citep[e.g.,][]{bell07, ilbert10, lefloch05, magnelli11, rodighiero10}. The question whether the joint evolution of MFs and IR LFs is self-consistent has received comparatively little attention, excepting studies confronting stellar MFs at $z$\,$\lesssim$\,3--5 with the integrated, higher-redshift star formation (SF) history \citep[e.g.,][]{wilkins08, leborgne09}. The common link between the stellar mass (\mstar) and SFR of galaxies is the specific star formation rate (sSFR). Star-forming galaxies (SFGs) at both high and low redshifts obey a tight relation (dispersion $<$0.3\,dex) -- the ``galaxy main sequence" -- according to which sSFR is a constant or slowly decreasing function of \mstar ~\citep[e.g.,][]{brinchmann04, daddi07, elbaz07, elbaz11, noeske07, damen09, pannella09, karim11}. SFGs on the main sequence are complemented by ``starbursting" galaxies \citep[a few percent of the total population; e.g.,][]{rodighiero11} with highly elevated sSFRs. Differences between the shape of the IR spectral energy distribution \citep[SED;][]{elbaz11} of these two kinds of systems and the different efficiency with which they convert molecular gas to stars \citep{daddi10b, genzel10} have given rise to the notion of ``bimodal" star formation.\\
The stellar MF of SFGs is well-fitted by a Schechter function \citep[e.g.,][]{bell07, ilbert10}, whereas the IR LF (used here as proxy for the SFR distribution) is generally parameterized as a double-exponential \citep[e.g.,][]{lefloch05} or double-power-law function \citep[e.g.,][]{magnelli09}. A possible astrophysical origin of this difference is the occurrence of burst-like and ``normal" (main-sequence-like) SF activity among SFGs. The evolving shape of IR LFs hence implicitly contains information on the relative importance of the two modes of SF in the past \citep[e.g.,][]{franceschini01, bethermin11}.\\
In this Letter, we show how the contribution of main-sequence and starburst galaxies to IR LFs at  $z$\,$\leq$\,2 can be predicted (Section \ref{sect:results}) using a simple scheme relying on basic observables (the evolution of sSFR in main-sequence galaxies and the sSFR distribution at fixed \mstar; see Section \ref{sect:setstage}) and starting from the evolution of the stellar MF of SFGs.

\section{(s)SFR and \mstar; basic characterization of star-forming galaxies}
\label{sect:setstage}

\subsection{Stellar Mass Functions and Main-sequence Evolution at $z$\,$<$\,2}
\label{sect:MFnMS}

\begin{figure}
\epsscale{1.09}
\centering
\plotone{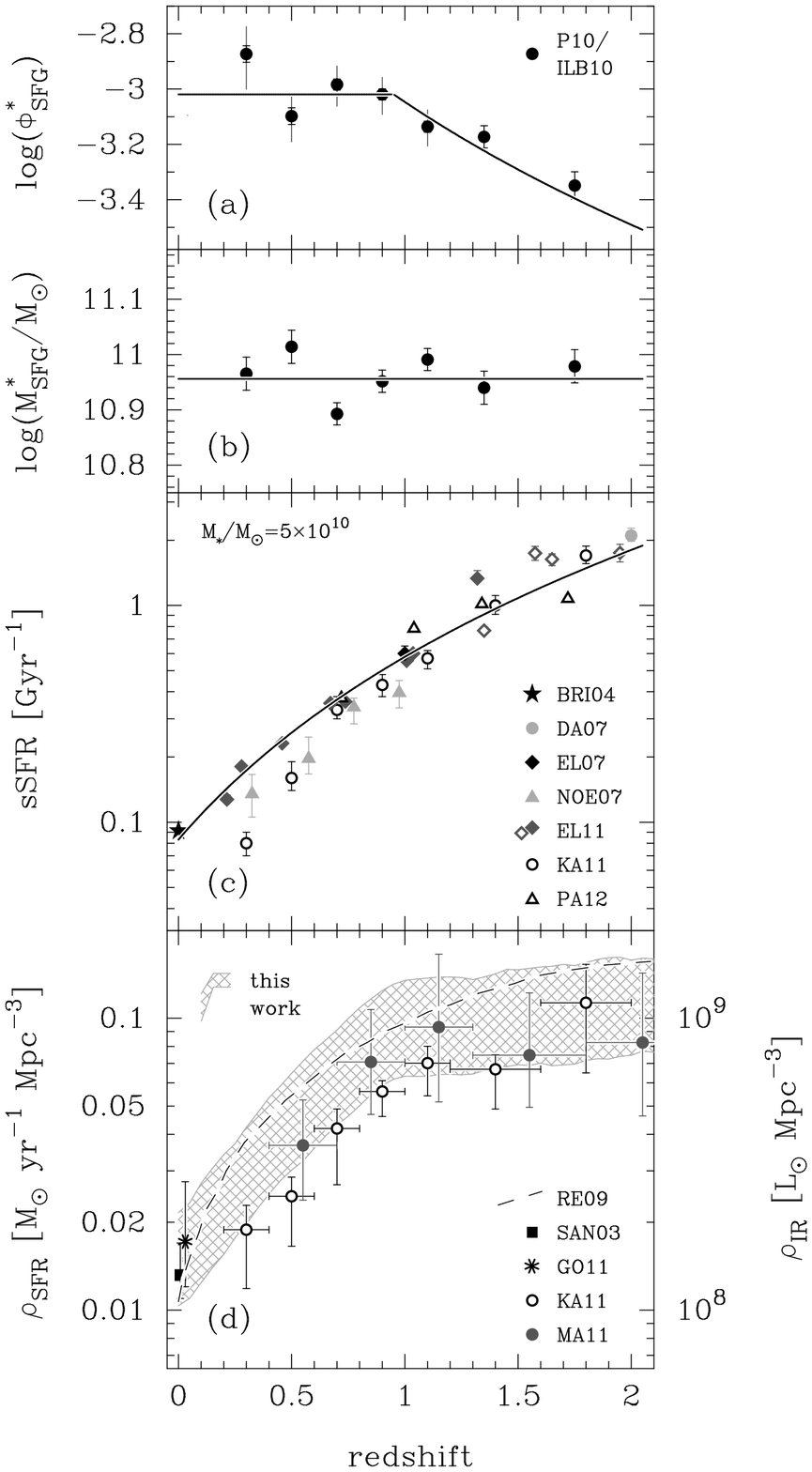}
\caption{Evolution of characteristic density {\it (a)} and mass {\it (b)} of the MF of star-forming galaxies, of the sSFR of main-sequence galaxies with $M_{\star}/M_{\odot}$\,=\,5$\times$10$^{10}$ {\it (c{\rm ; open symbols indicate stacking results})}, and of the cosmic SFRD at $z$\,$<$\,2 {\it (d{\rm; shaded area shows evolution predicted in this analysis})}. Solid lines in panels {\it (a)}--{\it (c)} trace average evolutionary trends (Section \ref{sect:setstage}). In panel {\it (a)} grey error bars at $z$\,$\leq$\,1.1 denote cosmic variance estimates from \citet{scoville07}. (See references for key to authorship abbreviations.) \label{fig:ingred}}
\end{figure}

Schechter function fits to the stellar MF of SFGs vary with sample selection \citep[e.g. the color cut adopted to separate blue from red galaxies;][]{baldry11} and the comparison of different measurements is additionally complicated by covariance among Schechter parameters. Nevertheless, the general consensus in the literature is that, for SFGs, the low-mass end slope $\alpha$ and characteristic mass $M^*$ change little at 0\,$<$\,$z$\,$<$\,2, while the normalization $\Phi^*$ is roughly constant out to $z$\,$\sim$\,1, followed by a $\sim$3-fold drop over 1\,$<$\,$z$\,$<$\,2 \citep[e.g.,][]{bell07, pozzetti10, brammer11}.\\
In the following we will use the fits by \citet{peng10} of the MFs in \citet{ilbert10} which include both sources with intermediate and high SF activity (i.e. should encompass the {\it entire} star-forming population). Assuming a fixed $\alpha$\,=\,-1.3, the error-weighted, mean characteristic mass at $z$\,$<$\,2 is  log($\langle M^*\rangle$/$M_{\odot}$)\,=\,10.96$\pm$0.01 (Fig. \ref{fig:ingred}.b\footnote{All data were converted to a WMAP-7 cosmology \citep{larson11} with ($H_0$\,[km\,s$^{-1}$\,Mpc$^{-1}$], $\Omega_{\rm m}$, $\Omega_{\Lambda}$) = (70.4, 0.273, 0.727) and a \citet{chabrier03} IMF.}). Density fluctuations in the COSMOS field at intermediate redshift (e.g. $z$\,$\sim$\,0.3) cause significant scatter among $\Phi^*$ measurements at $z$\,$<$\,1. For simplicity we adopt a constant average of log($\Phi^*$)\,=\,-3.02$\pm$0.01 which is consistent with all data below $z$\,$\sim$\,1 if both measurement and cosmic variance errors are accounted for (Fig. \ref{fig:ingred}.a). After $z$\,$\sim$\,1, $\Phi^*$ decreases as (1+$z$)$^{-2.40^{+0.22}_{-0.34}}$.

\noindent Fig. \ref{fig:ingred}.c shows the redshift evolution of the sSFR of SFGs with $M_{\star}/M_{\odot}$\,$\simeq$\,5$\times$10$^{10}$. This evolution is well documented in recent literature on the SF main sequence at $z$\,$\lesssim$\,2. Here we use results based on a combination of mid-IR and UV data \citep{daddi07, elbaz07, noeske07}, far-IR data \citep[][M. Pannella et al. 2012, in preparation]{elbaz11} and radio continuum imaging \citep{karim11}. Error bars denote the uncertainty on the average sSFR($z$) rather than intrinsic scatter (generally $\sim$0.3\,dex); they hence are often masked by the plotting symbols. The sSFR-measurements at $z$\,$>$\,0 define a tight trend that is well-fitted by a power law (1+$z$)$^{2.8\pm0.1}$ (equivalent to eq. 13 in \citealp{elbaz11}) and which, furthermore, accurately connects to the $z$\,=\,0 measurement of \citet[underlying data from \citealp{brinchmann04}]{elbaz07}. As a slight exception to the generally excellent agreement between the (1+$z$)$^{2.8}$-evolution and observations, the sSFR measurements of \citet{noeske07} are offset to lower sSFRs by about 1.5\,$\sigma$ -- possibly due to their shallow mid-IR data which might also be responsible for the steeper decline of sSFR with \mstar~these authors find -- but seem to follow the same evolutionary slope. The sample of \citet{karim11} is identical to the one used to compute the MF of panels {\it (a)} and {\it (b)}. As explicitly discussed by these authors, the evolution and slope/normalization of the sSFR versus \mstar~relation are selection-dependent; the lower sSFRs they find at $z$\,$\lesssim$\,0.5 thus reflect the growing importance of the ``intermediate"-activity population at low redshift.

\begin{figure*}
\centering
\includegraphics[scale=.6, angle=-90]{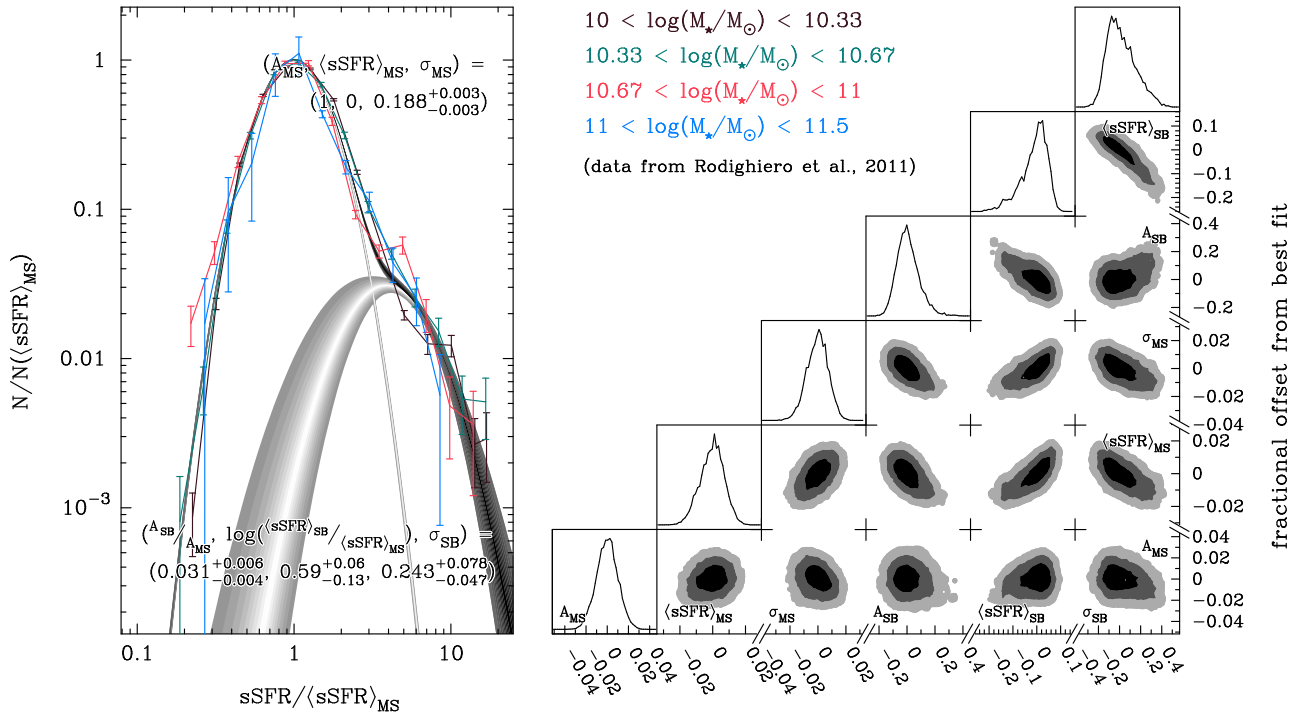}
\caption{{\it Left}: double-Gaussian decomposition (main-sequence, MS, and starburst, SB, activity) of the sSFR distribution at fixed \mstar\,$>$\,10$^{10}$\msun~for galaxies at $z$\,$\sim$\,2. sSFR-distributions are re-aligned as described in Section \ref{sect:DGdecomp}. Grey bands span 95\% confidence regions; white (black) cores trace the preferred main-sequence/starburst (total) distribution. Text inserts: best-fit parameters (median of the posterior probability distribution and 95\% confidence limits) for the normalized double-Gaussian plotted.\newline
{\it Right}: double-Gaussian parameter covariance; light grey, dark grey, and black areas extend to the 3, 2 and 1\,$\sigma$ confidence contours. Offsets are parametrized in fractions of the preferred value (X\,$\in$\,\{MS, SB\}): amplitudes $A_{\rm X}$ -- linear;  peak position $\langle{\rm sSFR}\rangle_X$ and width $\sigma_{\rm X}$ -- logarithmic. Panels along upper edge show the MCMC output distributions.\newline
{\it Note: in the journal Fig. \ref{fig:DGdecomp} presents a reduced version of this plot to comply with ApJL panel restrictions.}
\label{fig:DGdecomp}}
\end{figure*}

\subsection{Decomposition into Main-sequence and Starburst Contribution}
\label{sect:DGdecomp}

Using a combination of BzK- and {\it Herschel}/PACS-selected SFGs at 1.5\,$<$\,$z$\,$<$\,2.5 from the COSMOS and GOODS fields, \citet{rodighiero11} showed that the sSFR-distribution of galaxies at fixed \mstar~tapers out into a broad tail of starbursting galaxies beyond the SF main sequence (sSFRs of these starbursts may exceed main-sequence values more than 10-fold). By imposing a hard sSFR-cut above which SFGs are considered ``starbursts", \citet{rodighiero11} estimated that merely 2\% of massive ($M_{\star}/M_{\odot}$\,$\geq$\,10$^{10}$) galaxies have a burst-like nature and contribute $\lesssim$10\% to the SFR-density (SFRD). We now decompose the sSFR distributions of \citet{rodighiero11} into two Gaussian components (subsequently identified as main-sequence and burst-like SF activity, respectively):
\begin{eqnarray}
&N&{\rm (sSFR)}|_{M_{\star}} = \\ \nonumber
&&\sum_{\rm X\in\{MS,\,SB\}} A_{\rm X}\,{\rm exp}\left(-\frac{({\rm sSFR}-\langle{\rm sSFR}\rangle_{\rm X})^2}{2\sigma_{\rm X}^2}\right)~. \label{eq:DG}
\end{eqnarray}
The physical motivation behind this approach is that, e.g., galaxy interactions need not strongly boost SF \citep[e.g.][]{dimatteo07}, and that -- at fixed \mstar~-- it thus appears more natural if the importance of burst-like SF activity successively grows with increasing (s)SFR. We interpret the double-Gaussian shape as follows: individual SFGs are ``hybrid" objects harboring both main-sequence and burst-mode SF activity in relative proportions that vary according to their sSFR. SFGs well offset from the locus where the main-sequence and starburst Gaussian have equal amplitudes are increasingly dominated by only one mode of SF.

\noindent In their analysis, \citet[][see their Fig. 2]{rodighiero11} find that the distributions of sSFR at 1.5\,$<$\,$z$\,$<$\,2.5 in four stellar mass bins spanning 10\,$<$\,log($M_{\star}/M_{\odot}$)\,$<$\,11.5 are self-similar, albeit displaced in sSFR according to $M_{\star}^{-0.21\pm0.04}$ (implying a slope of $\sim$0.79 for the main sequence in the SFR versus $M_{\star}$ plane) and with different amplitudes that reflect the shape of the MF. This self-similarity yields much better constraints on the free parameters of the double-Gaussian decomposition than could be obtained by decomposing stellar mass bins individually. Having re-scaled the data of \citet{rodighiero11} to a common reference frame (see Fig. \ref{fig:DGdecomp}, left), we estimated the six free parameters describing our double-Gaussian distribution with a Monte Carlo Markov Chain (MCMC; 10$^6$ realizations). We provide best-fit parameters of a normalized\footnote{The direct fit for, e.g., galaxies at $z$\,=\,2 with 10.33\,$<$\,log(\mstar/\msun)\,$<$\,10.66 gives ($A_{\rm SB}$, $\langle {\rm sSFR}\rangle_{\rm SB}$, $\sigma_{\rm SB}$) = (2.2$^{+0.4}_{-0.3}$$\times$$10^{-5}$\,Mpc$^{-3}$\,dex$^{-1}$, 0.893$^{+0.063}_{-0.131}$\,Gyr$^{-1}$, 0.243$^{+0.078}_{-0.047}$\,dex) and ($A_{\rm MS}$, $\langle {\rm sSFR}\rangle_{\rm MS}$, $\sigma_{\rm MS}$) = (70.0$\pm$1.4, 0.303$^{+0.004}_{-0.005}$, 0.188$\pm$0.003) in identical units. Errors quoted are 95\% confidence limits.} distribution (amplitude/position of the starburst Gaussian are expressed relative to the main-sequence Gaussian) in Fig. \ref{fig:DGdecomp}. The posterior probability distributions of the individual parameters are well behaved (viz., unimodal) but subject to some covariance, especially among the parameters of the starburst component (cf. Fig. \ref{fig:DGdecomp}, right).

\noindent With our double-Gaussian decomposition of the (s)SFR distribution, we obtain a modified estimate of the contribution of starburst activity to the SFRD 
at $z$\,$\sim$\,2 of 14.2$^{+1.7}_{-1.3}$\% (68\% confidence limits) as opposed to \citet{rodighiero11} who find $\sim$10\% when considering only sources with sSFR/$\langle {\rm sSFR} \rangle_{\rm MS}$\,$>$\,4. We do not update the computation of the number density of starbursts in \citet{rodighiero11} because galaxies below their sSFR-threshold are likely hybrid sources where normal and burst-like SF coexist.

\begin{figure*}
\centering
\includegraphics[scale=.94]{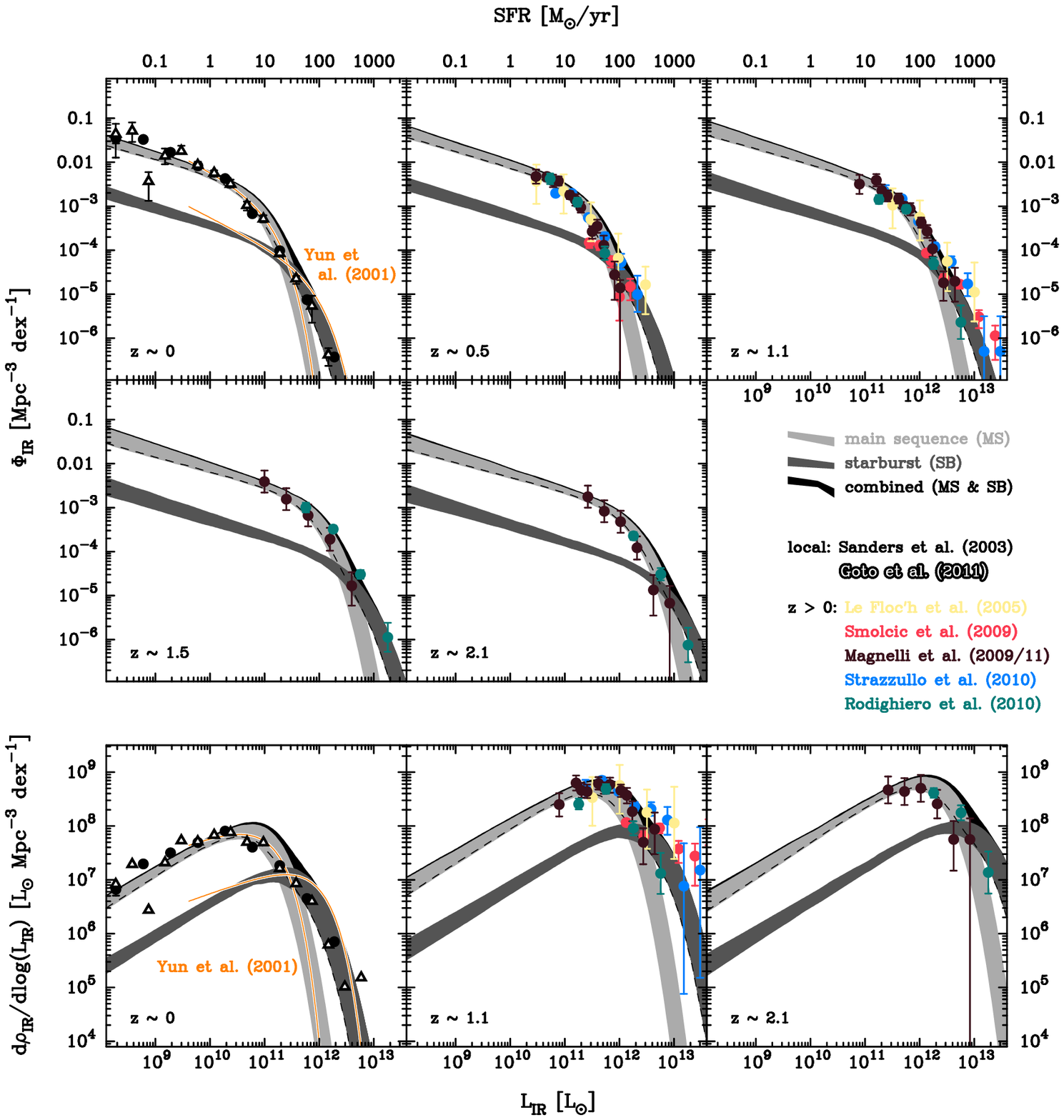}
\caption{{\it Top}: predicted contribution of normal ({\it light grey}) and burst-like ({\it dark grey}) star formation to IR LFs (SFR distributions; conversion between SFR and \lir~following \citealp{kennicutt98}) at $z$\,$\lesssim$\,2. Shaded areas: 68\% confidence region resulting from uncertainties on the evolution of the stellar MF, the cosmic evolution of sSFR and the double-Gaussian decomposition of Section \ref{sect:DGdecomp}. Overlaid literature measurements (see legend and additional explanations in text) match the predictions well. At $z$\,$\sim$\,0 we also plot the double-Schechter decomposition of \citet[{\it orange line}]{yun01}.\newline
{\it Bottom}: predicted IR luminosity density distributions. \label{fig:IR-LFs}}
\end{figure*}

\section{Results}
\label{sect:results}

To construct IR (8-1000\,$\mu$m) LFs for the interval $z$\,$\leq$\,2 we make two assumptions: (1) the double-Gaussian decomposition of Fig. \ref{fig:DGdecomp}, performed for galaxies at $z$\,$\sim$\,2, remains valid {\it at all $z$\,$<$\,2, implying an unchanged contribution to the SFRD} of normal and burst-like SF activity; (2) the slope of the main sequence (measured to be $\sim$0.79) and the double-Gaussian decomposition {\it do not change at masses below those studied by \citet{rodighiero11}}. The comparison of predicted IR LFs with observations will reveal whether these simplifications are justified. We also assume that {\it low-sSFR outliers to the main sequence do not contribute significantly} \citep[e.g.][]{salmi12} to the IR LF. Although little is known about the low-sSFR tail of the distributions in Fig. \ref{fig:DGdecomp}, this simplification seems justified as most of these sources are passive galaxies undergoing little obscured SF.\\
The mapping of the stellar MF to an IR LF is effectively a convolution of the MF and a variable double-Gaussian kernel with (1) normalization fixed  by the shape of the MF and (2) main-sequence peak position that -- given the redshift -- is uniquely determined by the position of the SF main sequence in the (s)SFR versus \mstar~plane. Thanks to our starburst vs. main-sequence decomposition, we can then also identify the individual contribution of normal and burst-like SF activity to the IR LF.

\begin{figure}[h!]
\epsscale{1.18}
\plotone{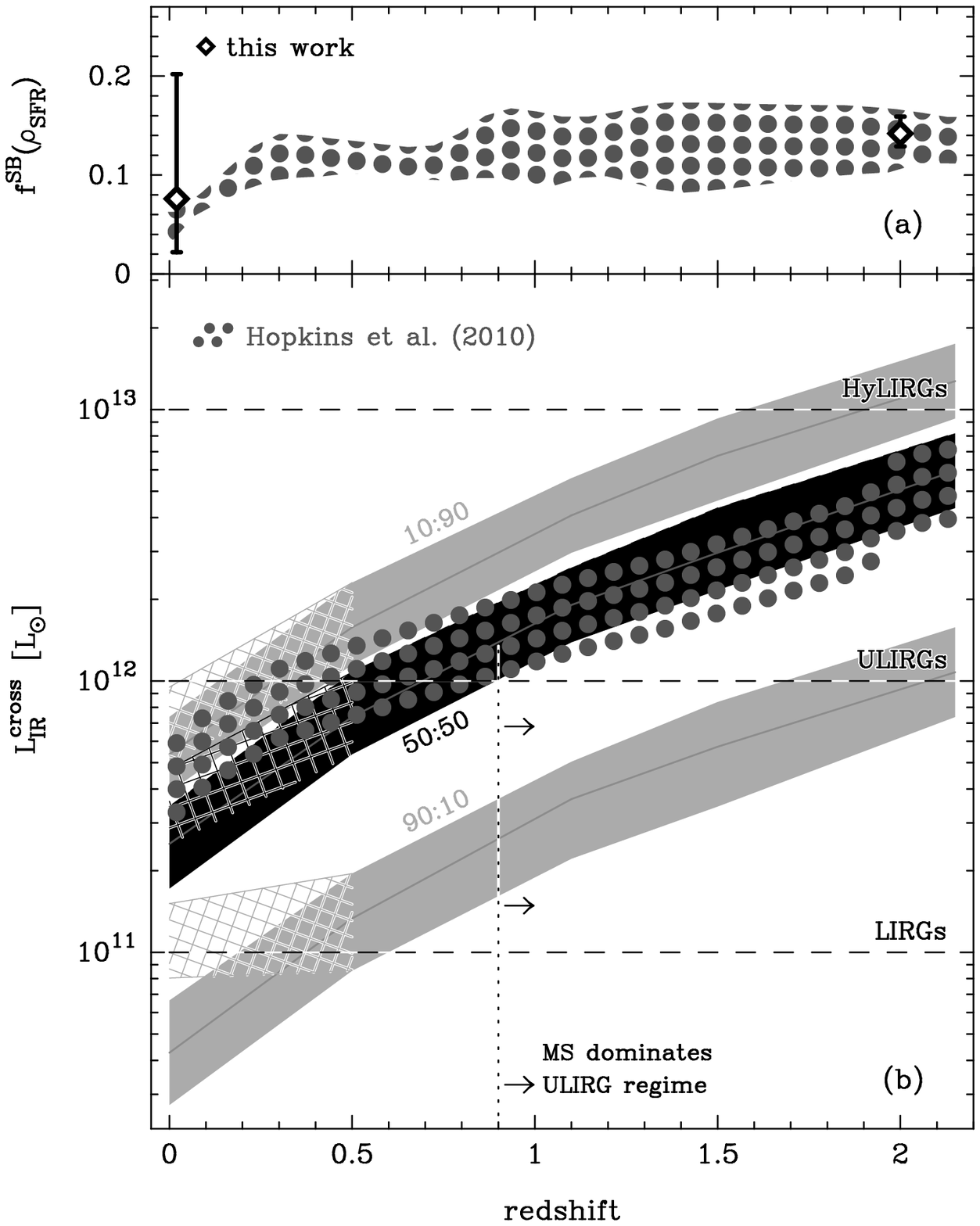}
\caption{{\it (a)}: Constraints on the fractional contribution $f^{\rm SB}(\rho_{\rm SFR})$ of starbursts to the cosmic SFRD at $z$\,=\,0 and 2, compared to simulation-based evolutionary predictions in \citet{hopkins10} (dotted area).
\newline
{\it (b)}: Redshift evolution of the cross-over luminosity for equal contributions of normal and burst-like star formation to IR LFs ({\it black}; 68\% confidence region highlighted). Light grey curves: luminosities at which main-sequence activity contributes 90\% (10\%) to the global IR LF. Hatched branches: low-redshift evolution assuming the best-fit $f^{\rm SB}(\rho_{\rm SFR}$,\,$z$\,=\,0)\,=\,7.6$^{+12.6}_{-5.4}$\% of panel {\it (a)}. Main-sequence star formation dominates the ULIRG regime at $z$\,$\gtrsim$\,0.9 (dotted vertical line).
\label{fig:LIRthreshevo}}
\end{figure}

\noindent Fig. \ref{fig:IR-LFs} shows our predicted IR LFs (see also Table \ref{tab:LFtab}) and luminosity density distributions for five redshift bins spanning 0\,$<$\,$z$\,$<$\,2.1. The starburst component (dark grey) has a reduced amplitude and is shifted to higher luminosities than the main-sequence contribution (light grey). We compare our prediction with measurements of the $z$\,$>$\,0 IR LFs of \citet{lefloch05} and \citet{rodighiero10}, both based on {\it Spitzer}/MIPS 24\,$\mu$m photometry. We also plot the LF of \citet{magnelli11} who combine MIPS 24 and 70\,$\mu$m data\footnote{The {\it Herschel}/PACS-based IR LFs in \citet{gruppioni10} are consistent with {\it Spitzer} studies. We refrain from overplotting their LFs because they do not help constrain the starburst contribution.}. The 1.4\,GHz LFs for (color-selected) SFGs in \citet{smolcic09} and \citet{strazzullo10} have been converted to $L_{\rm IR}$ by applying the IR-radio relation which holds out to at least $z$\,$\sim$\,2 \citep[see][for a corresponding analysis conducted on the sample of \citealp{smolcic09}]{sargent10}. These observations closely agree with our predicted IR LF. At high luminosities a marginally significant excess is observed for some data points from \citet{lefloch05}, \citet{smolcic09} and \citet{strazzullo10}, likely caused by residual contamination from active galactic nucleus (AGN) related processes that affect mid-IR and radio continuum fluxes more than far-IR emission. Note that our ``model" IR LFs deliberately {\it include} the contribution to the SFRD from {\it SF activity in AGN hosts}.\\
We emphasize that our formalism is ``anchored" to $z$\,$\sim$\,2 where the double-Gaussian decomposition is directly constrained by the data of \citet{rodighiero11}. A more stringent test is whether the local IR LF can be correctly reproduced if our framework is left to evolve to $z$\,=\,0. In this respect, the agreement of our predicted $z$\,$\sim$\,0 IR LF with the local IR LFs of \citet{sanders03} and \citet{goto11} far up the faint-end slope is encouraging and indicates that the simplifying assumptions noted at the beginning of this section (i.e. invariance of the decomposition with redshift and stellar mass) are legitimate. At the faint end, our LF rises as $\nicefrac{dN}{dL_{\rm IR}}$\,$\propto$\,\lir$^{-1.4}$, in broad agreement with, e.g., the recent measurement of the faint-end slope (Schechter $\alpha$\,=\,-1.51$\pm$0.08) of the SFR distribution of \citet{bothwell11}. The exact $\alpha$ we predict depends on the slope of the sSFR-\mstar~relation and the inclusion of low-sSFR outliers (neglected here) to the main sequence could also cause some additional steepening. At bright \lir~our split of the $z$\,$\sim$\,0 LF matches an earlier double-Schechter decomposition of the local {\it IRAS} LF into a ``spiral galaxy" and starburst component by \citet[][plotted in orange in Fig. \ref{fig:IR-LFs} over the luminosity range for which their decomposition fits their {\it IRAS} data]{yun01}.\\
Finally we plot (Fig. \ref{fig:ingred}.d; hatched area) our evolutionary prediction for the SFR-density ($\rho_{\rm SFR}$) -- obtained by integrating the LFs of Fig. \ref{fig:IR-LFs} -- which is consistent with literature data. The cosmic SF history has been traced by countless studies, of which we show two recent analyses relying on extinction-free tracers of SFR (\citealp{karim11}, radio continuum; \citealp{magnelli11}, far-IR emission), plus the evolution inferred from extinction-corrected UV data in \citet{reddy09}.

\section{Discussion}
\label{sect:discussion}

By performing a double-Gaussian fit (Fig. \ref{fig:DGdecomp}) to the distribution of (s)SFR at fixed \mstar~reported in \citet{rodighiero11}, we have re-estimated the starburst contribution (14.2$^{+1.7}_{-1.3}$\%) to the SFRD at $z$\,$\sim$\,2. We then introduce a self-consistent framework which successfully predicts the evolution of IR LFs at all $z$\,$<$\,2. This approach improves over the conceptually similar analysis of \citet{bell07} by quantifying the relative importance of main-sequence and burst-like SF since $z$\,$\sim$\,2. The bimodal nature of our LF model -- tabulated in Table \ref{tab:LFtab} for reference -- defines a cross-over IR-luminosity ($L_{\rm IR}^{\rm cross}$) above which starburst activity dominates over ``quiescent" (main-sequence) SF activity. In Fig. \ref{fig:LIRthreshevo}.b we illustrate the evolution of $L_{\rm IR}^{\rm cross}$ from log(\lir/\lsun)\,=\,11.4 to 12.8 at  0\,$<$\,$z$\,$<$\,2, as well as similar thresholds (separated by approx. one decade of \lir) where main-sequence and starburst activity contributes 90\% (10\%) to the total IR LF. These thresholds evolve in parallel by a factor (1+$z$)$^{2.8}$ if we assume a redshift-invariant starburst fraction and a constant slope of the main sequence in the (s)SFR versus \mstar~plane, i.e. in our framework the luminosity evolution of the SFR-distribution reflects the cosmic evolution of sSFR in main-sequence galaxies. Similarly, the density scale at $L_{\rm IR}^{\rm cross}$ mirrors the density evolution of the MF of SFGs.\\
We reproduce the well-known fact that most local ULIRGs are starbursts \citep[e.g.,][]{sandersmirabel96}. At $z$\,$\gtrsim$\,0.9 the majority of ULIRGs are main-sequence galaxies. Importantly, however, their high SFR ($>$100\,\msun) is not triggered by merging as in most local ULIRGs but is a secular process linked to large gas reservoirs in these high-redshift disks \citep[e.g.,][]{daddi10a, tacconi10}. Local and distant ULIRGs are {\it intrinsically different objects} for which direct comparisons should be avoided; the selection of galaxy populations by \lir~is problematic since any conclusion will depend on the redshift-range considered.

\noindent It is natural to link starbursts to merging activity even though not all high-sSFR outliers display unmistakable signs of galaxy-galaxy interactions \citep[e.g.,][]{kartaltepe11}. The fact that the evolution of the IR LF is compatible with a universal starburst fraction is surprising since merger rates are expected to decrease with cosmic time \citep[e.g.,][]{kitzbichler08}. Some cosmological hydrodynamic simulations \citep[e.g.,][]{hopkins10} predict an increase of the merger-induced, burst-like contribution to the SFRD from a few percent locally to $\sim$12\% (Fig. \ref{fig:LIRthreshevo}.a), in apparent contradiction to the approach adopted here. If one leaves the relative amplitudes of starburst and main-sequence Gaussian distributions free to vary in the fit of the local IR LF, a starburst contribution $f^{\rm SB}(\rho_{\rm SFR})$\,=\,7.6$^{+12.6}_{-5.4}$\% (68\% confidence limits) to the local SFRD is derived. This value lies between the prediction of \citet{hopkins10} and the observationally estimated 15\%-20\% in, e.g., \citet{brinchmann04} or \citet{kennicutt05} and is also consistent with non-evolution out to $z$\,=\,2. By drawing random pairs of $f^{\rm SB}(\rho_{\rm SFR})$ at $z$\,=\,0 and 2 from within the error distributions of our measurements we were nevertheless able to determine that there is a 75\% probability that $f^{\rm SB}(\rho_{\rm SFR})$ increases out to $z$\,=\,2. Generally speaking, the good match between our predicted and the observed LFs suggests that variations of starburst fraction with mass, as well as the neglected contribution of low-sSFR outliers to the main sequence do not strongly influence the shape of the IR LF.\\
The success of this simple picture at $z$\,$\leq$\,2 motivates an extension to higher redshift using the known evolution of the sSFR \citep[e.g.][]{gonzalez10} and provides a powerful framework for the prediction of cosmological observables, e.g. the evolution of molecular gas reservoirs in normal and starburst galaxies (M.~T. Sargent et al. 2012, in preparation) or source counts in the IR, making use of the distinctly different SED for main-sequence and starburst galaxies (M. B\'ethermin et al. 2012, in preparation).

\acknowledgments
We thank E. Le Floc'h, G. Lagache, and H. Aussel for helpful discussions. J. Mullaney, M. Pannella, G. Rodighiero, V. Strazzullo, and an anonymous referee provided feedback and/or data. M.T.S., M.B. and E.D. were supported by grants ERC-StG UPGAL 240039 and ANR-08-JCJC-0008.\\

\begin{deluxetable}{lccc}
\tabletypesize{\small}
\tablewidth{11truecm}
\tablecaption{Predicted IR LFs at $z$\,$\sim$\,1 (68\% confidence limits; cf. Fig. \ref{fig:IR-LFs}). \label{tab:LFtab}}
\tablehead{
\colhead{log(\nicefrac{\lir}{\lsun})} &
\colhead{log($\Phi_{\rm tot}$ [Mpc$^{-3}$\,dex$^{-1}$])} &
\colhead{log($\Phi_{\rm MS}$)} &
\colhead{log($\Phi_{\rm SB}$)}}
\startdata
8.10 & -1.22$_{-0.15}^{+0.16}$ & -1.26$_{-0.14}^{+0.17}$ & -2.40$_{-0.17}^{+0.18}$\\
8.20 & -1.26$_{-0.15}^{+0.16}$ & -1.30$_{-0.14}^{+0.16}$ & -2.44$_{-0.17}^{+0.18}$\\
8.30 & -1.30$_{-0.14}^{+0.15}$ & -1.33$_{-0.14}^{+0.15}$ & -2.47$_{-0.17}^{+0.17}$\\
8.40 & -1.34$_{-0.14}^{+0.15}$ & -1.36$_{-0.14}^{+0.14}$ & -2.51$_{-0.17}^{+0.16}$\\
8.50 & -1.37$_{-0.13}^{+0.14}$ & -1.40$_{-0.14}^{+0.14}$ & -2.55$_{-0.16}^{+0.16}$\\
8.60 & -1.41$_{-0.13}^{+0.14}$ & -1.43$_{-0.14}^{+0.13}$ & -2.59$_{-0.15}^{+0.16}$\\
8.70 & -1.45$_{-0.13}^{+0.13}$ & -1.47$_{-0.14}^{+0.12}$ & -2.63$_{-0.15}^{+0.16}$\\
8.80 & -1.48$_{-0.13}^{+0.13}$ & -1.51$_{-0.13}^{+0.12}$ & -2.67$_{-0.14}^{+0.16}$\\
8.90 & -1.52$_{-0.12}^{+0.12}$ & -1.55$_{-0.12}^{+0.12}$ & -2.70$_{-0.14}^{+0.15}$\\
9.00 & -1.56$_{-0.12}^{+0.12}$ & -1.59$_{-0.12}^{+0.12}$ & -2.74$_{-0.14}^{+0.15}$\\
9.10 & -1.60$_{-0.12}^{+0.11}$ & -1.63$_{-0.12}^{+0.12}$ & -2.78$_{-0.14}^{+0.15}$\\
9.20 & -1.63$_{-0.12}^{+0.11}$ & -1.67$_{-0.12}^{+0.11}$ & -2.82$_{-0.13}^{+0.14}$\\
9.30 & -1.67$_{-0.12}^{+0.10}$ & -1.71$_{-0.11}^{+0.11}$ & -2.85$_{-0.13}^{+0.14}$\\
9.40 & -1.71$_{-0.12}^{+0.10}$ & -1.75$_{-0.11}^{+0.11}$ & -2.89$_{-0.13}^{+0.14}$\\
9.50 & -1.75$_{-0.12}^{+0.10}$ & -1.79$_{-0.11}^{+0.11}$ & -2.93$_{-0.12}^{+0.14}$\\
9.60 & -1.79$_{-0.11}^{+0.10}$ & -1.82$_{-0.11}^{+0.11}$ & -2.97$_{-0.12}^{+0.14}$\\
9.70 & -1.83$_{-0.11}^{+0.10}$ & -1.86$_{-0.10}^{+0.10}$ & -3.01$_{-0.11}^{+0.13}$\\
9.80 & -1.87$_{-0.10}^{+0.10}$ & -1.90$_{-0.09}^{+0.10}$ & -3.05$_{-0.12}^{+0.13}$\\
9.90 & -1.90$_{-0.10}^{+0.09}$ & -1.94$_{-0.09}^{+0.09}$ & -3.08$_{-0.11}^{+0.12}$\\
10.00 & -1.94$_{-0.10}^{+0.09}$ & -1.98$_{-0.09}^{+0.09}$ & -3.12$_{-0.11}^{+0.11}$\\
10.10 & -1.98$_{-0.09}^{+0.08}$ & -2.01$_{-0.08}^{+0.09}$ & -3.15$_{-0.11}^{+0.11}$\\
10.20 & -2.02$_{-0.08}^{+0.08}$ & -2.05$_{-0.08}^{+0.09}$ & -3.19$_{-0.11}^{+0.11}$\\
10.30 & -2.06$_{-0.09}^{+0.08}$ & -2.09$_{-0.09}^{+0.09}$ & -3.23$_{-0.10}^{+0.10}$\\
10.40 & -2.10$_{-0.09}^{+0.08}$ & -2.13$_{-0.09}^{+0.09}$ & -3.27$_{-0.09}^{+0.10}$\\
10.50 & -2.15$_{-0.08}^{+0.08}$ & -2.18$_{-0.08}^{+0.08}$ & -3.31$_{-0.09}^{+0.11}$\\
10.60 & -2.18$_{-0.08}^{+0.07}$ & -2.22$_{-0.08}^{+0.08}$ & -3.35$_{-0.09}^{+0.11}$\\
10.70 & -2.22$_{-0.08}^{+0.07}$ & -2.26$_{-0.08}^{+0.08}$ & -3.39$_{-0.09}^{+0.11}$\\
10.80 & -2.27$_{-0.08}^{+0.08}$ & -2.30$_{-0.08}^{+0.08}$ & -3.43$_{-0.09}^{+0.10}$\\
10.90 & -2.32$_{-0.08}^{+0.08}$ & -2.36$_{-0.08}^{+0.09}$ & -3.47$_{-0.09}^{+0.11}$\\
11.00 & -2.38$_{-0.09}^{+0.09}$ & -2.42$_{-0.08}^{+0.10}$ & -3.51$_{-0.10}^{+0.10}$\\
11.10 & -2.44$_{-0.10}^{+0.10}$ & -2.48$_{-0.10}^{+0.11}$ & -3.55$_{-0.09}^{+0.10}$\\
11.20 & -2.51$_{-0.10}^{+0.11}$ & -2.54$_{-0.11}^{+0.10}$ & -3.60$_{-0.09}^{+0.10}$\\
11.30 & -2.57$_{-0.13}^{+0.11}$ & -2.61$_{-0.13}^{+0.12}$ & -3.64$_{-0.09}^{+0.10}$\\
11.40 & -2.65$_{-0.14}^{+0.13}$ & -2.70$_{-0.14}^{+0.13}$ & -3.69$_{-0.10}^{+0.10}$\\
11.50 & -2.75$_{-0.16}^{+0.15}$ & -2.80$_{-0.15}^{+0.16}$ & -3.73$_{-0.10}^{+0.09}$\\
11.60 & -2.86$_{-0.18}^{+0.16}$ & -2.92$_{-0.19}^{+0.18}$ & -3.79$_{-0.11}^{+0.08}$\\
11.70 & -3.00$_{-0.19}^{+0.21}$ & -3.06$_{-0.22}^{+0.22}$ & -3.85$_{-0.11}^{+0.08}$\\
11.80 & -3.15$_{-0.23}^{+0.24}$ & -3.23$_{-0.25}^{+0.26}$ & -3.91$_{-0.12}^{+0.08}$\\
11.90 & -3.33$_{-0.26}^{+0.30}$ & -3.43$_{-0.30}^{+0.33}$ & -3.99$_{-0.13}^{+0.09}$\\
12.00 & -3.52$_{-0.29}^{+0.33}$ & -3.65$_{-0.35}^{+0.38}$ & -4.08$_{-0.16}^{+0.12}$\\
12.10 & -3.73$_{-0.32}^{+0.36}$ & -3.91$_{-0.43}^{+0.43}$ & -4.17$_{-0.18}^{+0.14}$\\
12.20 & -3.96$_{-0.33}^{+0.38}$ & -4.22$_{-0.50}^{+0.50}$ & -4.29$_{-0.19}^{+0.17}$\\
12.30 & -4.19$_{-0.32}^{+0.39}$ & -4.58$_{-0.58}^{+0.57}$ & -4.42$_{-0.22}^{+0.20}$\\
12.40 & -4.43$_{-0.32}^{+0.39}$ & -5.00$_{-0.66}^{+0.65}$ & -4.57$_{-0.24}^{+0.24}$\\
12.50 & -4.67$_{-0.32}^{+0.39}$ & -5.48$_{-0.75}^{+0.74}$ & -4.74$_{-0.28}^{+0.28}$\\
12.60 & -4.89$_{-0.35}^{+0.38}$ & -6.02$_{-0.85}^{+0.81}$ & -4.94$_{-0.32}^{+0.31}$\\
12.70 & -5.14$_{-0.37}^{+0.39}$ & -6.62$_{-0.95}^{+0.89}$ & -5.15$_{-0.36}^{+0.34}$\\
12.80 & -5.40$_{-0.42}^{+0.41}$ & -7.30$_{-1.06}^{+0.98}$ & -5.41$_{-0.42}^{+0.39}$\\
12.90 & -5.71$_{-0.45}^{+0.46}$ & -8.05$_{-1.18}^{+1.07}$ & -5.71$_{-0.45}^{+0.45}$\\
13.00 & -6.04$_{-0.51}^{+0.51}$ & -8.87$_{-1.31}^{+1.16}$ & -6.04$_{-0.51}^{+0.51}$\\
13.10 & -6.39$_{-0.61}^{+0.55}$ & -9.79$_{-1.42}^{+1.28}$ & -6.39$_{-0.61}^{+0.55}$\\
13.20 & -6.76$_{-0.71}^{+0.57}$ & -10.78$_{-1.56}^{+1.40}$ & -6.76$_{-0.71}^{+0.57}$\\
13.30 & -7.22$_{-0.80}^{+0.64}$ & -11.85$_{-1.70}^{+1.52}$ & -7.22$_{-0.80}^{+0.64}$\\
13.40 & -7.67$_{-0.90}^{+0.68}$ & -13.01$_{-1.84}^{+1.62}$ & -7.67$_{-0.90}^{+0.68}$\\
13.50 & -8.16$_{-1.02}^{+0.69}$ & -14.28$_{-1.96}^{+1.75}$ & -8.16$_{-1.02}^{+0.69}$\\
13.60 & -8.75$_{-1.15}^{+0.80}$ & -15.65$_{-2.08}^{+1.88}$ & -8.75$_{-1.15}^{+0.80}$\\
\enddata
\tablecomments{~MS: main sequence; SB: starburst; tot: MS+SB. LFs at $z$\,$\neq$\,1 follow by applying a luminosity scaling (1+$z$)$^{2.8}$ with density fixed for $z$\,$<$\,1 and with density varying according to $\Phi(z)$\,$\propto$\,(1+$z$)$^{-2.4}$ at 1\,$<$\,$z$\,$<$\,2.}
\end{deluxetable}


\begin{thebibliography}{}
\bibitem[Baldry et al.(2011)]{baldry11} Baldry, I.~K., Driver, S.~P., Loveday, J., et al.\ 2011, arXiv:1111.5707
\bibitem[Bell et al.(2003)]{bell03} Bell, E.~F., McIntosh, D.~H., Katz, N., \& Weinberg, M.~D.\ 2003, \apjl, 585, L117
\bibitem[Bell et al.(2007)]{bell07} Bell, E.~F., Zheng, X.~Z., Papovich, C., et al.\ 2007, \apj, 663, 834
\bibitem[B{\'e}thermin et al.(2011)]{bethermin11} B{\'e}thermin, M., Dole, H., Lagache, G., Le Borgne, D., \& Penin, A.\ 2011, \aap, 529, A4
\bibitem[Bothwell et al.(2011)]{bothwell11} Bothwell, M.~S., Kenicutt, R.~C., Johnson, B.~D., et al.\ 2011, \mnras, 415, 1815
\bibitem[Brammer et al.(2011)]{brammer11} Brammer, G.~B., Whitaker, K.~E., van Dokkum, P.~G., et al.\ 2011, \apj, 739, 24
\bibitem[Brinchmann et al.(2004)]{brinchmann04} Brinchmann, J., Charlot, S., White, S.~D.~M., et al.\ 2004, \mnras, 351, 1151 (BRI04)
\bibitem[Chabrier(2003)]{chabrier03} Chabrier, G.\ 2003, \pasp, 115, 763
\bibitem[Daddi et al.(2007)]{daddi07} Daddi, E., et al.\ 2007, \apj, 670, 156 (DA07)
\bibitem[Daddi et al.(2010a)]{daddi10a} Daddi, E., et al.\ 2010a, \apj, 713, 686
\bibitem[Daddi et al.(2010b)]{daddi10b} Daddi, E., et al.\ 2010b, \apjl, 714, L118
\bibitem[Damen et al.(2009)]{damen09} Damen, M., Labb{\'e}, I., Franx, M., et al.\ 2009, \apj, 690, 937
\bibitem[Di Matteo et al.(2007)]{dimatteo07} Di Matteo, P., Combes, F., Melchior, A.-L., \& Semelin, B.\ 2007, \aap, 468, 61
\bibitem[Elbaz et al.(2007)]{elbaz07} Elbaz, D., et al.\ 2007, \aap, 468, 33 (EL07)
\bibitem[Elbaz et al.(2011)]{elbaz11} Elbaz, D., Dickinson, M., Hwang, H.~S., et al.\ 2011, \aap, 533, A119 (EL11)
\bibitem[Franceschini et al.(2001)]{franceschini01} Franceschini, A., Aussel, H., Cesarsky, C.~J., Elbaz, D., \& Fadda, D.\ 2001, \aap, 378, 1
\bibitem[Genzel et al.(2010)]{genzel10} Genzel, R., et al.\ 2010, \mnras, 407, 2091
\bibitem[Gonz{\'a}lez et al.(2010)]{gonzalez10} Gonz{\'a}lez, V., Labb{\'e}, I., Bouwens, R.~J., et al.\ 2010, \apj, 713, 115
\bibitem[Goto et al.(2011)]{goto11} Goto, T., Arnouts, S., Inami, H., et al.\ 2011, \mnras, 410, 573 (GO11)
\bibitem[Gruppioni et al.(2010)]{gruppioni10} Gruppioni, C., Pozzi, F., Andreani, P., et al.\ 2010, \aap, 518, L27 
\bibitem[Hopkins et al.(2010)]{hopkins10} Hopkins, P.~F., Younger, J.~D., Hayward, C.~C., Narayanan, D., \& Hernquist, L.\ 2010, \mnras, 402, 1693
\bibitem[Ilbert et al.(2010)]{ilbert10} Ilbert, O., Salvato, M., Le Floc'h, E., et al.\ 2010, \apj, 709, 644 (ILB10)
\bibitem[Karim et al.(2011)]{karim11} Karim, A., et al.\ 2011, \apj, 730, 61 (KA11)
\bibitem[Kartaltepe et al.(2011)]{kartaltepe11} Kartaltepe, J.~S., Dickinson, M., Alexander, D.~M., et al.\ 2011, arXiv:1110.4057
\bibitem[Kennicutt(1998)]{kennicutt98} Kennicutt, R.~C., Jr.\ 1998, \araa, 36, 189
\bibitem[Kennicutt et al.(2005)]{kennicutt05} Kennicutt, R.~C., Lee, J.~C., Akiyama, S., Funes, J.~G., \& Sakai, S.\ 2005, in AIP Conf. Proc. 783, The Evolution of Starbursts (Melville, NY: AIP), 3
\bibitem[Kitzbichler \& White(2008)]{kitzbichler08} Kitzbichler, M.~G., \& White, S.~D.~M.\ 2008, \mnras, 391, 1489
\bibitem[Larson et al.(2011)]{larson11} Larson, D., Dunkley, J., Hinshaw, G., et al.\ 2011, \apjs, 192, 16
\bibitem[Le Borgne et al.(2009)]{leborgne09} Le Borgne, D., Elbaz, D., Ocvirk, P., \& Pichon, C.\ 2009, \aap, 504, 727
\bibitem[Le Floc'h et al.(2005)]{lefloch05} Le Floc'h, E., et al.\ 2005, \apj, 632, 169
\bibitem[Magnelli et al.(2009)]{magnelli09} Magnelli, B., Elbaz, D., Chary, R.~R., Dickinson, M., Le Borgne, D., Frayer, D.~T., \& Willmer, C.~N.~A.\ 2009, \aap, 496, 57
\bibitem[Magnelli et al.(2011)]{magnelli11} Magnelli, B., Elbaz, D., Chary, R.~R., Dickinson, M., Le Borgne, D., Frayer, D.~T., \& Willmer, C.~N.~A.\ 2011, \aap, 528, A35 (MA11)
\bibitem[Noeske et al.(2007)]{noeske07} Noeske, K.~G., Weiner, B.~J., Faber, S.~M., et al.\ 2007, \apjl, 660, L43 (NOE07)
\bibitem[Pannella et al.(2009)]{pannella09} Pannella, M., Carilli, C.~L., Daddi, E., et al.\ 2009, \apjl, 698, L116
\bibitem[Pannella et al.(2012)]{pannella12} Pannella, M., et al.\ 2012, in prep. (PA12)
\bibitem[Peng et al.(2010)]{peng10} Peng, Y.-J., Lilly, S.~J., Kova{\v c}, K., et al.\ 2010, \apj, 721, 193 (P10)
\bibitem[Pozzetti et al.(2010)]{pozzetti10} Pozzetti, L., Bolzonella, M., Zucca, E., et al.\ 2010, \aap, 523, A13
\bibitem[Reddy \& Steidel(2009)]{reddy09} Reddy, N.~A., \& Steidel, C.~C.\ 2009, \apj, 692, 778  (RE09)
\bibitem[Rodighiero et al.(2010)]{rodighiero10} Rodighiero, G., et al.\ 2010, \aap, 515, A8
\bibitem[Rodighiero et al.(2011)]{rodighiero11} Rodighiero, G., Daddi, E., Baronchelli, I., et al.\ 2011, \apjl, 739, L40
\bibitem[Salmi et al.(2012)]{salmi12} Salmi, F., Daddi, E., Elbaz, D., et al.\ 2012, ApJ, submitted 
\bibitem[Sanders \& Mirabel(1996)]{sandersmirabel96} Sanders, D.~B., \& Mirabel, I.~F.\ 1996, \araa, 34, 749 
\bibitem[Sanders et al.(2003)]{sanders03} Sanders, D.~B., Mazzarella, J.~M., Kim, D.-C., Surace, J.~A., \& Soifer, B.~T.\ 2003, \aj, 126, 1607 (SAN03)
\bibitem[Sargent et al.(2010)]{sargent10} Sargent, M.~T., Schinnerer, E., Murphy, E., et al.\ 2010, \apjl, 714, L190
\bibitem[Scoville et al.(2007)]{scoville07} Scoville, N., Aussel, H., Benson, A., et al.\ 2007, \apjs, 172, 150
\bibitem[Smol{\v c}i{\'c} et al.(2009)]{smolcic09} Smol{\v c}i{\'c}, V., et al.\ 2009, \apj, 690, 610
\bibitem[Strazzullo et al.(2010)]{strazzullo10} Strazzullo, V., Pannella, M., Owen, F.~N., et al.\ 2010, \apj, 714, 1305 
\bibitem[Tacconi et al.(2010)]{tacconi10} Tacconi, L.~J., et al.\ 2010, \nat, 463, 781
\bibitem[Wilkins et al.(2008)]{wilkins08} Wilkins, S.~M., Trentham, N., \& Hopkins, A.~M.\ 2008, \mnras, 385, 687
\bibitem[Yun et al.(2001)]{yun01} Yun, M.~S., Reddy, N.~A., \& Condon, J.~J.\ 2001, \apj, 554, 80
\end{thebibliography}
\end{document}